\documentclass[twoside,runningheads]{llncs}

\usepackage{graphicx}

\usepackage{amsmath}

\usepackage{cite} %

\usepackage{hyperref}
\usepackage{xcolor}

\usepackage[irlabel]{fcps}

\usepackage{prettyref}
\newcommand{\rref}[2][]{\prettyref{#2}}
\newrefformat{sec}{Section\,\ref{#1}}
\newrefformat{def}{Def.\,\ref{#1}}
\newrefformat{thm}{Theorem\,\ref{#1}}
\newrefformat{prop}{Proposition\,\ref{#1}}
\newrefformat{lem}{Lemma\,\ref{#1}}
\newrefformat{cor}{Corollary\,\ref{#1}}
\newrefformat{ex}{Example\,\ref{#1}}
\newrefformat{tab}{Table\,\ref{#1}}
\newrefformat{fig}{Fig.\,\ref{#1}}
\newrefformat{app}{Appendix\,\ref{#1}}

\usepackage{ifpdf}
\ifpdf
\fi

\begin{document}

\title{The Significance of Symbolic Logic for\newline Scientific Education\thanks{%
Funding has been provided by an Alexander von Humboldt Professorship.}}
\titlerunning{The Significance of Symbolic Logic for Scientific Education}
\author{Andr\'e Platzer\orcidID{0000-0001-7238-5710}}
\institute{
  Karlsruhe Institute of Technology, Karlsruhe, Germany
  \email{platzer@kit.edu}
}

\maketitle

\begin{abstract}
This invited paper is a passionate pitch for the significance of logic in scientific education.
Logic helps focus on the essential core to identify the foundations of ideas and provides corresponding longevity with the resulting approach to new and old problems.
Logic operates symbolically, where each part has a precise meaning and the meaning of the whole is compositional, so a simple function of the meaning of the pieces.
This compositionality in the meaning of logical operators is the basis for compositionality in reasoning about logical operators.
Both semantic and deductive compositionalities help explain what happens in reasoning.
The correctness-critical core of an idea or an algorithm is often expressible eloquently and particularly concisely in logic.
The opinions voiced in this paper are influenced by the author's teaching of courses on cyber-physical systems, constructive logic, compiler design, programming language semantics, and imperative programming principles.
In each of those courses, different aspects of logic come up for different purposes to elucidate significant ideas particularly clearly.
While there is a bias of the thoughts in this paper toward computer science, some courses have been heavily frequented by students from other majors so that some transfer of the thoughts to other science and engineering disciplines is plausible.
\end{abstract}

\section{Introduction}

This paper is a passionate pitch for the significance of logic in scientific education.
Education has multiple important goals that range from practical skills that enable a student to efficiently solve present challenges met in academic and industrial practice all the way to the longevity of foundations that influence a student's thinking for a lifetime in yet unforeseeable areas.
Logic is particularly good at impacting the latter foundations but also plays a role in the former practice.
The reason for logic's longevity is its unreasonable effectiveness \cite{Wigner60} of identifying the core essentials of a question and its answer.
For example, several aspects that make a big syntactic difference in a particular application context still lead to a negligible difference once the phenomenon has been captured with logic.
Just like proof-based mathematics, its sibling of logic teaches rigorous reasoning principles and how to use it to overcome challenges.
Logic is crystal clear on the foundations of reasoning, on what a proof is, which reasoning principles are correct and why, and is explicit about possible structuring principles about proofs.
There is an obvious and frequently mentioned positive synergy with experience in proof-based mathematics and experience in logic.

The opinions reflected in this paper are based on the author's experience in computer science undergraduate to PhD-level teaching at different research universities since 2008 as well as engagement in university committees restructuring undergraduate introductory education.
While these opinions are influenced from the author's educational computer science bias, the introductory Principles of Imperative Computation and the upper-level Logical Foundations of Cyber-Physical Systems courses he taught were also frequented by students from several other majors, including mathematics, physics, electrical engineering, and robotics, so that some generalization of these thoughts to other sciences and engineering disciplines are plausible.

\section{Logic in the Science Curriculum}

To give the reader some appreciations for the different aspects of logic that enter different courses, here is an overview of some of the logic-influences in some of the courses influenced by what the author taught.\footnote{%
Lecture material is available at \url{http://lfcps.org/}
}
\begin{enumerate}
\item Principles of Imperative Computation is the first computer science course for computer science undergraduates and many other disciplines originally designed by Frank Pfenning and refined by the author. It combines algorithms and data structures, imperative programming introduction, and program contracts for establishing their correctness. Logic enters informally when reasoning mathematically about preconditions, postconditions, loop invariants, and data structure invariants. This immediate appreciation for the fundamental aspects of operational and logical reasoning about programs enables the students in the course to obtain a particularly strong command of writing reliable imperative programs. In the last homework, the students showcase their understanding of imperative programming by programming a virtual machine in C reflecting on the nuances of operational semantics.
About 300--800 students take this course every year.
\label{item:PIC}

\item Constructive Logic is a third year undergraduate course for computer science but also mathematics and philosophy students designed by Frank Pfenning, also taught by Karl Crary, and refined by the author. It teaches the logical foundations of functional programming from proofs-as-programs via the Curry-Howard isomorphism as well as the logical foundations of logic programming from propositions-as-programs while proving the fundamental principle of cut elimination and proof search refinements along the way.
About 100 students take this course every year.
\label{item:CLogic}

\item Compiler Design is an upper undergraduate level computer science course designed by Frank Pfenning and the author. It teaches the principles for designing compilers, where logical proof rules that represent logic programs are used to capture dataflow analysis, parsing, programming language semantics (via structural operational semantics), compiler optimizations etc.
While logic is not a learning goal of the course, logic is still crucial in concisely communicating the actual idea behind many phases of a compiler.
About 80 students take this course every year.
\label{item:compilers}

\item Logical Foundations of Cyber-physical Systems \cite{DBLP:conf/cpsed/Platzer13,Platzer18} is an upper undergraduate and/or master's level computer science course designed by the author. It teaches what principles one has to know in order to design safe cyber-physical systems and justify their safety properties. Logic and programming languages play central roles in the course as means to identify the core mathematical challenges and solution techniques especially for justifying safety. The course is supported by a textbook, slides, lecture videos, a theorem prover, active learning quizzes, and a competition.
About 20 students take this course every year.
\label{item:LFCPS}

\item Programming Language Semantics is a PhD-level course based on Steve Brooke's take on John Reynolds course design \cite{Reynolds}, but reinterpreted by the author with more focus on logic. It teaches the principles of programming language semantics, which has a direct link to logic via the axiomatic semantics of programming languages, as well as clear links with logic via soundness and relative completeness theorems of the axiomatic semantics with respect to the denotational or operational semantics. The extensive use of logic in the author's version of the course significantly simplifies otherwise quite demanding and technical semantical challenges making it possible to teach crucial ideas with minimal technical effort.
About 10--20 students take this course every year.
\label{item:PLS}
\end{enumerate}

Some of the insights behind some of these courses with a particular focus on logic will be reviewed in the sequel.
The course in items~\ref{item:PIC} is a required first semester undergraduate course.
The courses in items~\ref{item:CLogic} and~\ref{item:LFCPS} are one of several different courses satisfying the undergraduate logic requirement.
The course in item~\ref{item:compilers} is one of several different courses satisfying the systems requirement.
The courses in items~\ref{item:LFCPS}  and~\ref{item:PLS} are one of several different courses satisfying the programming language PhD requirement.

\section{Course: Logical Foundations of Cyber-Physical Systems}

The \emph{Logical Foundations of Cyber-Physical Systems} (LFCPS) course\footnote{\url{http://www.lfcps.org/lfcps/}} is accompanied by a textbook \cite{Platzer18}, slides, more than 20 hours of lecture videos\footnote{\url{http://videos.lfcps.org/}}, and a thorough active learning quiz that enables students to practice and reinforce the learning goals of every lecture.
The course was originally designed for upper level undergraduate students \cite{DBLP:conf/cpsed/Platzer13} but has since been opened up to masters and PhD students.
It has been designed and taught by the author at Carnegie Mellon University, ENS Lyon, the University of Braga, and the Karlsruhe Institute of Technology.
Short versions of the course were the basis for several summer school lectures, including the summer schools in Marktoberdorf, on Automated Reasoning, on Verification Technology Systems \& Applications, and on Cyber-Physical Systems.
The following discussion is based on thoughts expressed in the LFCPS textbook \cite{Platzer18}, which also identifies lecture dependencies for courses based on different subsets of the lectures.

The Logical Foundations of Cyber-Physical Systems textbook and course are breaking with the myth that cyber-physical systems (CPSs) are too challenging to be taught at the undergraduate level. CPSs such as computer-controlled cars, airplanes or robots play an increasingly crucial role in our daily lives. They are systems that we bet our lives on, so they need to be safe. Getting CPSs safe, however, is an intellectual challenge, because of their intricate interactions of complex control software with their physical behavior. Who can design these notoriously challenging systems with the scrutiny that is required to make sure they can be used safely? How can students, scientists, and practitioners acquire the required background in a single course or a single textbook in a way that meets the demands on rigor required in safe CPS design?
In the LFCPS course, students quickly advance from learning basic concepts underlying CPSs to being able to prove safety properties about complex CPS.

\paragraph{The Challenge.}
Teaching CPS-related topics is notoriously challenging, but also creates an opportunity to discover and explore other areas of science with the intrinsic motivation it takes to succeed.  A few students may have a background either in engineering physical systems, or in some areas of formal methods, but almost never in both and, in fact, often in neither of the two. A sharp educational gap has also been confirmed across the board at the 2013 NSF workshop on CPS education \cite{DBLP:conf/cpsed/2013}. This brings up the question of how to best teach the core aspects of CPS with the rigor that is required to prepare students and professionals for the challenges that lie ahead in enriching our world with safe and reliable cyber-physical systems.

The challenge is that CPSs are a cozy topic to take on after background reading equivalent to the material acquired for a PhD in mathematics, a PhD in computer science, a PhD in logic, and a PhD in engineering or controls.
The trick is to find out how to enable students to understand CPS gradually without losing interest and engagement.
Besides identifying the easiest and most intuitive, background-free approach for presenting the various technical concepts, the biggest educational contribution of this course is the clever out-of-order arrangement of the topics to overcome the problem that CPSs have such a long dependency chain of required background.
There certainly is a reason for the acute shortage of rigorous CPS courses for undergraduate students.

There are two primary ways of learning about cyber-physical systems \cite{Platzer18}, reprinted here with permission from Springer:

\paragraph{Onion Model.}
The \emph{Onion Model} follows the natural dependencies of the layers of mathematics going outside in, peeling off one layer at a time, and progressing to the next layer when all prerequisites have been covered.
This would require the CPS student to first study all relevant parts of computer science, mathematics, and engineering, and then return to CPS in the big finale.
That would require the first part of the course to cover real analysis, the second part differential equations, the third part conventional discrete programming, the fourth part classical discrete logic, the fifth part theorem proving, and finally the last part cyber-physical systems.
In addition to the significant learning perseverance that the Onion Model requires, a downside is that it misses out on the integrative effects of CPSs that can bring different areas of science and engineering together, and which provide a unifying motivation for studying them in the first place.

\paragraph{Scenic Tour Model.}
The LFCPS course follows the \emph{Scenic Tour Model}, which starts at the heart of the matter, namely CPSs, going on scenic expeditions in various directions to explore the world around as we find the need to understand the respective subject matter.
The course directly targets CPS right away, beginning with simpler layers that the reader can understand in full before moving on to the next challenge.

For example, the first layer comprises CPSs without feedback control, which allow simple finite open-loop controls to be designed, analyzed, and verified without the technical challenges considered in later layers of CPS.
Likewise, the treatment of CPS is first limited to cases where the dynamics can still be solved in closed form, such as straight-line accelerated motion of Newtonian dynamics, before generalizing to systems with more challenging differential equations that can no longer be solved explicitly.
This gradual development where each level is mastered and understood and practiced in full before moving to the next level is helpful to tame complexity.
It also follows naturally the layers of complexity in logic.
The Scenic Tour Model has the advantage that the students stay on CPSs the whole time, and leverage them as the guiding motivation for understanding more and more about the connected areas.
It has the disadvantage that the resulting gradual development of CPS does not necessarily always present matters in the same way that an after-the-fact compendium would treat it.\footnote{%
The textbook compensates for this by providing technical summaries and by highlighting important results for later reference.
}
A gradual development can also be more effective at conveying the ideas, reasons, and rationales behind the development compared to a final compendium, which improve generalizability promises compared to a mere factual presentation.

\paragraph{Computational Thinking for CPS.}
The approach that the LFCPS course follows takes advantage of Computational Thinking \cite{DBLP:journals/cacm/Wing06} for CPSs.
Due to their subtleties and the intricate interactions of complex control software with the physical world, CPSs are notoriously challenging.
Logical scrutiny, formalization, and thorough safety and correctness arguments are, thus, critical for CPS.
Because CPS are so easy to get wrong, these logical aspects are an integral part of their design and critical to understanding their complexities.

The primary attention of the course, thus, is on the foundations and core principles of CPS.
The course tames some of the CPS complexities by focusing on a simple core programming language for CPS.
The elements of the programming language are introduced hand in hand with their reasoning principles, which makes it possible to combine CPS program design with their safety arguments.
This is important, not just because abstraction is a key factor for success in CPS, but also because retrofitting safety is not possible in CPS.
The CPS programming language of hybrid programs taught in the textbook pass with flying colors Alan J. Perlis' test, who rejected programming languages by the following criterion:
\begin{quote}
``A language that doesn't affect the way you think about programming, is not worth knowing.''
\\\hfill~\makebox{--~Alan~J.~Perlis}~\cite{DBLP:journals/sigplan/Perlis82}
\end{quote}

\paragraph{Logic to Tame CPS.}
On account of their technical challenges, programming and investigating the safety of CPS may be a daunting task.
But logic significantly simplifies this challenge thanks to the principles of logical compositionality in multi-dynamical systems \cite{DBLP:conf/lics/Platzer12a,DBLP:conf/cade/Platzer16,Platzer18}.
The first step is to make safety statements about cyber-physical systems a first-level citizen in logic.
The resulting differential dynamic logic (\dL) \cite{DBLP:journals/jar/Platzer08,Platzer08,Platzer10,DBLP:conf/lics/Platzer12a,DBLP:journals/jar/Platzer17,Platzer18} features modalities \(\dbox{\asprg}{}\) for hybrid programs $\asprg$ that describe the possible behavior of a CPS as a program with differential equations.
The \dL formula \(\dbox{\asprg}{\asfml}\) expresses that after all runs of hybrid program $\asprg$ the \dL formula $\asfml$ is true (safety).
It is an ordinary \dL formula, so the implication \(\bsfml \limply \dbox{\asprg}{\asfml}\) expresses that if formula $\bsfml$ is true initially, then all runs of hybrid program $\asprg$ are such that formula $\asfml$ is true afterwards.

\begin{example}[Car acceleration or braking] \label{ex:car-accel-brake}
The following \dL formula expresses that, if a car $x$ is before an obstacle $m$ and its brakes $b$ work, then all ways of following a hybrid program that first has a nondeterministic choice ($\cup$) to apply acceleration by assigning \(\pupdate{\pumod{a}}{A}\) or to apply brake by \(\pupdate{\pumod{a}{-b}}\) and subsequently follows the differential equation system with the time-derivative $\D{x}$ of position $x$ being velocity $v$, whose time-derivative is the acceleration $a$ but only while the velocity $v$ is nonnegative, then all its behaviors keep the position before the obstacle and the velocity nonnegative:
\[
x\leq m \land b>0 \limply \dbox{\lpgroup\pchoice{\pupdate{\pumod{a}}{A}}{\pupdate{\pumod{a}}{-b}}\rpgroup;\lpbrace\pevolvein{\D{x}=v\syssep\D{v}=a}{v\geq0}\rpbrace}{(x\leq m \land 0\leq v)}
\]
Whether this \dL formula is true is a good question, but now this question has a logically precise rendition and unambiguous answer.
\end{example}

Logic is not just useful for the clear and unambiguous expression of questions about cyber-physical systems but also for finding the answer.
This is where axioms for cyber-physical systems become useful.
For example, the \dL axiom \irref{composeb} captures that all behavior of the sequence $\ausprg;\busprg$ safely satisfies formula $\ausfml$ if and only if all behaviors of the first part $\ausprg$ are such that all behaviors of the second part $\busprg$ satisfy $\ausfml$:
\[
\cinferenceRule[composeb|$\dibox{{;}}$]{composition} %
{\linferenceRule[equiv]
  {\dbox{\ausprg}{\dbox{\busprg}{\ausfml}}}
  {\axkey{\dbox{\ausprg;\busprg}{\ausfml}}}
}{}%
\]

\begin{example}[Car motion after car control] \label{ex:car-accel-brake-;}
Using axiom \irref{composeb} reduces the \dL formula from \rref{ex:car-accel-brake} to an equivalent that split off the discrete control from the differential equation of motion:
\[
x\leq m \land b>0 \limply \dbox{\pchoice{\pupdate{\pumod{a}}{A}}{\pupdate{\pumod{a}}{-b}}}{\dbox{\pevolvein{\D{x}=v\syssep\D{v}=a}{v\geq0}}{(x\leq m \land 0\leq v)}}
\]
The advantage of this equivalent decomposition is that it separates the discrete control actions from the continuous motion such that both can be analyzed separately.
Subsequent logical decompositions of the remaining parts in \dL will ultimately lead to a significantly easier equivalent.
\end{example}

The following \dL equivalence can be used to analyze or prove conjunctions in the safety conditions separately:
\[
\dinferenceRule[band|${[]\land}$]{$\dbox{\cdot}{\land}$}
{\linferenceRule[equiv]
  {\dbox{\ausprg}{\ausfml} \land \dbox{\ausprg}{\busfml}}
  {\axkey{\dbox{\ausprg}{(\ausfml\land\busfml)}}}
}{}%
\]

\begin{example}[Car safety and speed separated] \label{ex:car-accel-brake-&}
Using axiom \irref{band} reduces the \dL formula from \rref{ex:car-accel-brake-;} to an equivalent that separately considers the position safety of $x\leq m$ and the speed safety $v\geq0$:
\begin{align*}
x\leq m \land b>0 \limply\,& \big(\dbox{\pchoice{\pupdate{\pumod{a}}{A}}{\pupdate{\pumod{a}}{-b}}}{\dbox{\pevolvein{\D{x}=v\syssep\D{v}=a}{v\geq0}}{\,x\leq m}}\\
&\land \dbox{\pchoice{\pupdate{\pumod{a}}{A}}{\pupdate{\pumod{a}}{-b}}}{\dbox{\pevolvein{\D{x}=v\syssep\D{v}=a}{v\geq0}}{\,0\leq v}}\big)
\end{align*}
Indeed the last conjunct is fairly easy to establish, because the differential equation system \(\pevolvein{\D{x}=v\syssep\D{v}=a}{v\geq0}\) is limited to $v\geq0$, which trivially implies the postcondition $0\leq v$.
But the same cannot be said about the first conjunct with the postcondition $x\leq m$.
Indeed, after some more decompositions by logical equivalences, it will turn out to be false if the initial velocity of the car exceeds its braking capabilities compared to the distance to the obstacle $m$.
\end{example}

The big conceptual advantage of working with logic for cyber-physical systems is that one can start with an unambiguous question phrased in \dL and then subsequently transform it with logic such that every step along the way is easy and clearly correct and the final outcome is easier than the original question.
It is, indeed, in many ways due to the use of logic that the LFCPS course is successful in simplifying the otherwise overwhelming challenges of cyber-physical systems by reducing them to simpler pieces \cite{DBLP:conf/cade/Platzer16}.

\paragraph{Active Learning Quizzes.}

The LFCPS course features active learning quizzes for every chapter and lecture of the accompanying textbook \cite{Platzer18}.
Learning by doing is a crucial element of understanding material. The purpose of the course quizzes is to support the student's learning by giving them an opportunity to practice and get feedback on how well they have achieved a selection of some of the learning goals of the LFCPS course.
By observing which ones the student is unsure about, the quizzes can help identify which material they should review again.
Since students ultimately need a solid understanding of all aspects of CPS, this helps stay up to speed.

The most profound impact of student learning stems from the ways of thinking that is internalized so deeply that the student can produce them on the fly without having to look anything up.
Concepts that become part of their thinking will enable students to autonomously detect situations where they apply, instead of needing to rely on others to tell them which concept to apply in order to solve which problem.

While quizzes feature carefully paced introductory questions, they are also designed to challenge a student's understanding.
This gives them an opportunity to think through some of the more subtle aspects of CPS at their own pace before they face similar challenges in application contexts where challenges may become overwhelming.
By solving a sequence of such separate challenges, students become better at understanding nuances and internalize the way of thinking that is required to solve them.
A few of the quiz questions give students an opportunity to synthesize multiple individual concepts to solve a small joint challenge.
These questions exercise synthetic knowledge and enable students to form conceptual bridges between individual skills to identify what they need where.

For example, some of the quiz questions ask students to check their thinking on certain simple subskills, which are useful to acquire early to avoid confusions.
Other quiz questions may make them wonder how long differential equations evolve and what exactly a safety property of a hybrid system means.
These are fundamental questions about CPS models that they can answer using their semantics.
Yet other quiz questions ask students to put all their acquired skills together to design simple CPS controllers or criticize their designs before facing the challenges of real applications.
Discovering a problem in one's thinking in the small context of a quiz question is a great learning experience and prevents students from the major downstream effects of carrying a conceptual misunderstanding forward into later parts of the course.
Quiz questions make students confront the blank page syndrome in the small, where they are asked to creatively come up with answers to small questions on their own.
This experience is not easy but prepares students for when they face bigger challenges where they will creatively come up with answers to bigger questions.

Except for the summary and wrapper questions with free text answers that are included for the purpose of reflection and feedback, the LFCPS active learning quizzes are fully autograded by the theorem prover \KeYmaeraX \cite{DBLP:conf/cade/FultonMQVP15} that proves correctness of the student's answer automatically and giving some question-dependent feedback when this fails.
It only happened twice that a student provided a correct answer on the active learning quiz that \KeYmaeraX was unable to prove.
By having even thought about this, the student arguably learned more than what could ever be reflected in the 2 missed points out of 500 quiz points.

\paragraph{Example Quiz Questions.}
The easiest quizzes to design are multiple-choice quizzes.
But those only teach students the passive skill of recognizing the right answer rather than the active skill of creatively producing the right answer.
This is essentially the educational counterpart to the computational P-NP problem \cite{DBLP:conf/stoc/Cook71}.
Checking correct answers is easier than producing correct answers.
One cannot learn integration by multiple-choice, because the mere process of differentiating the given answers masks the intended ability to learn how to integrate functions.
To give the reader a feeling how the LFCPS active learning quizzes feature genuinely \emph{active} learning, here is a small sampling of typical question types.

\begin{example}[Quiz: Program shapes] \label{ex:program-shapes}
            Objective (\emph{programming languages for CPS, semantics, models, operational effects}):
            {It is crucial to obtain an intuitive reading of the respective transitions in a hybrid program.
                This question gives you an opportunity to practice the mapping between a transition structure and the hybrid programs they correspond to.}

                What hybrid program fits to the following transition structure?

                \centerline{\includegraphics[width=5cm]{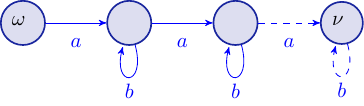}}

                Answer: \verb"{{a;{b;}*}*"\\
                Similar to an ellipsis, the dashed part of the
                transition systems indicates the transitions as shown that already happened
                before may happen again and again.
                Every \texttt{a} can be followed by an arbitrary
                number of \texttt{b}, as briefly indicated by the self loop with $b$, and the
                whole pattern can repeat arbitrarily often, as indicated by the dashed parts.
                Come to think of it, if the \texttt{b} transition goes back with a self-loop
                literally to the exact same state, then it cannot have had a huge effect or any
                at all. But, instead, the diagram illustrates the structure of the transition
                so the self-loops indicate that \texttt{b} can happen repeatedly at the
                indicated nodes, so \verb"{a;b;}*" would be incorrect, because it does not
                capture the fact that there can be an arbitrary number of \texttt{b} (even 0)
                between two consecutive \texttt{a} occurrences. The answer
                \verb"a;{b;}*;a;{b;}*;{a;{b;}*}*" would fit to the above transition
                structure as well, but is unnecessarily complicated, and, thus, not an
                insightful answer.

            For each of the following transition structures, find a hybrid program that can
            mimic its decisions. Give the simplest (shortest) hybrid program that can mimic
            all actions in the transition structure.
            
            \centerline{\includegraphics[width=5cm]{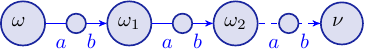}}

\end{example}

The quiz question type in \rref{ex:program-shapes} inverts definitions, giving students an opportunity to understand constructions on a deeper level by asking what input to a semantics definition would have a shape that fits to the expected output.
While the forward application of definitions is a crucial skill, the backward or inverse application requires reflection and practices deeper understandings of the inner working principles.
Without techniques for computing integrals, finding a function that has a given derivative requires a much deeper understanding about the process of differentiation.

\begin{example}[Quiz: True formulas] \label{ex:characterize-dL-truth}
Objective (\emph{model semantics, preconditions, rigorous specification}):
                {If a formula is not valid, it is important to identify when exactly it is true.
                    This helps identify missing preconditions to make it valid, and read off consequences when a formula is available as an assumption.
                    Of course, knowing when exactly a formula is true is also crucial when they are used as evolution domain constraints or tests, which is why those are usually quantifier-free FOL formulas.}
                    \\
                    Question: When is \dL formula \({\dbox{\pupdate{\pumod{x}{x+1}}}{\,x>5}}\) true?\\
                Answer: \texttt{x>4}\\
                Indeed, the first formula \({\dbox{\pupdate{\pumod{x}{x+1}}}{\,x>5}}\)
                and the second formula \(x>4\) are equivalent, i.e., they are true in exactly the same states, because the value of $x$ after the assignment \(\pupdate{\pumod{x}{x+1}}\) is one larger than it was before. So $x>5$ after \(\pupdate{\pumod{x}{x+1}}\) iff $x>4$ initially.
                But the latter formula $x>4$ is easier to understand than its equivalent \({\dbox{\pupdate{\pumod{x}{x+1}}}{\,x>5}}\), because it merely involves arithmetic evaluation, rather than also effectful programs.
                    
                    For each of the following \dL formulas identify the \textbf{exact} set of all
            states in which it is true and characterize this set by a quantifier-free
            formula of real arithmetic (of the same free variables). 
            \[\dbox{\pevolve{\D{x}=v\syssep\D{v}=a}}{\,x\leq m}\]
\end{example}

The quiz question type in \rref{ex:characterize-dL-truth} asks students to creatively come up with the answer when they learn something about continuous dynamics to characterize when exactly a car always stays before position $m$.
The answer is not easy but gives students a chance to practice and check their physical intuition as a logical formula.
Answering it helps ultimately fix and prove \rref{ex:car-accel-brake} in a subsequent synthetic skill question in a later quiz.
While answering this quiz question, students are led down a corridor of exploration of increasingly refined understandings of the safety of continuous dynamics.

\begin{example}[Quiz: Axiom usage] \label{ex:axiom-usage}
        Objective (\emph{rigorous reasoning about CPS}):
        {As one important part of rigorous reasoning about CPS, you will practice the correct application of axioms to differential dynamic logic problems.
          While the \KeYmaeraX prover correctly applies axioms for you, it is still helpful if you practice this yourself to get a better intuition for how it works and predict the outcome of a proof step before trying it.
          That will make you more time-efficient in your reasoning.
          It will also inform you how to transform parts of a proof to make useful axioms applicable later.
          If you properly understand reasoning principles, you are also better able to identify and check clever problem decompositions.}
        \\
        Question: What is the result of using axiom \irref{composeb} on \(\dbox{\text{ctrl};\text{plant}}{\,x>y}\)?\\
        Answer: \texttt{[ctrl;][plant;]x>y}
        \\
        Question: What is the result of using axiom \irref{assignb} on \(\dbox{\text{ctrl};\text{plant}}{\,x>y}\)?\\
        Answer: \texttt{n/a}\\
        Because the axiom \irref{assignb} expects an assignment instead of a $;$ as the top-level operator in the box modality, so is not applicable to the given formula, which is not of the form expected by the left-hand side of axiom \irref{assignb}.

      For each of the following \dL formulas, give the result of using the indicated \dL axiom (in its usual left to right decomposition direction) \textbf{once} to the whole formula \textbf{at its top-level position}, i.e., for the top-level operator and not deep down in the middle of some subformula.
      Respond \texttt{n/a} if the axiom is not applicable to the given formula.

      Axiom \irref{composeb} on \(\dbox{\text{sense};\text{ctrl};\text{plant}}{\,x>y}\)
\end{example}

The quiz question type in \rref{ex:axiom-usage} teaches students the precision of reasoning by asking them to predict a proof step of a theorem prover.
One might think that this skill is irrelevant, because a theorem prover such as \KeYmaeraX \cite{DBLP:conf/cade/FultonMQVP15} implementing differential dynamic logic will be perfectly capable of applying proof principles correctly.
But it is also helpful to develop an intuition for what will happen when, be able to predict the outcome, appreciate the nuances of reasoning overcome by a theorem prover along the way, and, especially, be able to do and check manual paper-based proofs as well.
Along the way of answering the above quiz question, students will discover on their own the significance of precision in the syntactic representation of programs.

\begin{example}[Quiz example: Axiom development] \label{ex:dL-axiom-development}
        Objective (\emph{operational CPS effects, \dL as a verification language}):
        {The axioms of differential dynamic logic are complete, so you do not need any more for its operators.
          But whenever you add new syntax to the language, then you give that operator a semantics, and also need to add new axioms for reasoning about the new syntactic features.
          These questions give you an opportunity to practice the extension of syntax, semantics, and axiomatics that fit together in harmony and properly decompose hybrid programs into logic.
          Recall that it is imperative that only sound axioms be adopted.
          Also remember that solid axioms for a program statement reduce the new syntactic program operator to simpler logical formulas about subformulas and subprograms, because that makes it possible to understand the new program operator solely in terms of easier logic, not in terms of different or more complicated programs.
          This question gives you the opportunity to practice the development of new axioms for hybrid programs.
        }

      The if-then-else statement \(\pif{\ivr}{\asprg}{\bsprg}\) runs HP $\asprg$ if formula $\ivr$ is initially true and runs HP $\bsprg$ otherwise.
      Its semantics is $\ll\dots\text{elided}\dots\gg$.

      Develop an axiom for \(\dbox{\pif{\ivr}{\ausprgax}{\busprgax}}{\ausfml}\)
      that decomposes the effect of the if-then-else statement in logic with simpler logical connectives.

\end{example}

After the axiom reflection question \rref{ex:axiom-usage}, and subsequent questions that ask students to check the correctness of conjectured axioms, the quiz question type in \rref{ex:dL-axiom-development} asks students to create rather than use axioms for reasoning about CPS, which enables them to develop higher metacritical analytic skills.

\begin{example}[Quiz: Loop invariants] \label{ex:loop-invariant-search}
      Objective (\emph{identifying and expressing invariants}):
      {The most important ingredient of a CPS is its invariant, because an invariant tells you what you always know about your system, no matter how long it operates.
        This question allows you to practice the important but challenging task of identifying loop invariants for hybrid systems.}

      Identify a loop invariant $\inv$ proving the following \dL formulas (after using rule \irref{implyr}) with exactly the following version of the loop invariant proof rule:

      \[
        \cinferenceRule[loop|loop]{inductive invariant}
        {\linferenceRule[sequent]
        {\lsequent{\Gamma} {\inv,\Delta}
          \quad\lsequent{\inv} {\dbox{\ausprg}{\inv}}
          \quad\lsequent{\inv} {\ausfml}}
        {\lsequent{\Gamma}{\dbox{\prepeat{\ausprg}}{\ausfml},\Delta}}
        }{}
      \]

      Write \texttt{n/a} when no loop invariant exists that proves the given \dL formula.

      \(x\geq1 \land v>0 \land A>0 \limply \dbox{\prepeat{\big\lpgroup
          \lpgroup\pchoice{\pupdate{\pumod{a}{0}}}{\pupdate{\pumod{a}{A}}}\rpgroup;
          \lpbrace\pevolve{\D{x}=v\syssep\D{v}=a}\rpbrace\big\rpgroup}}{\,x\geq0}\)

\end{example}

The quiz question type in \rref{ex:loop-invariant-search} implicitly follows up on the skills developed in \rref{ex:characterize-dL-truth} and gives students the opportunity to figure out why a control loop works correctly.
They are not kept guessing whether their answer is correct, because the \KeYmaeraX theorem prover underlying the active learning quizzes tells them right away when the proof did not succeed.
Other questions practice reflection and understanding by asking for the shortest possible invariant of a controller or differential equation, which helps students understand which arguments are necessary compared to which arguments are true but useless.
Subsequent quiz questions practice synthetic skills for system design and safety justifications while helping students understand the beneficial role that incremental system designs play in reducing verification and comprehension challenges.

\paragraph{CPS V\&V Grand Prix.}
In many but not all years, the LFCPS course also features practical verification labs culminating in self-defined course projects.
The students design, analyze, and prove correct a sequence of designs for robot models mastering increasingly difficult challenges. In each lab, the students design a controller for a single robot that can interact with an unknown environment. The students also design an appropriate model for the continuous behavior that their controlled robot would exhibit given the discrete control inputs. They need to decide on an appropriate model for the robot’s environment, including using nondeterminism to capture unknown behaviors in the environment. And finally, the students formalize a safety property as a logical formula and prove that their controller never violates it. The labs are all related and build on each other, with the ultimate goal that the students design and prove safety for a robot that can avoid moving obstacles.

Before students submit the final robot model (called \emph{Veribot}), they submit a \emph{Betabot}, which is a beta-version of the robot controller that they conjecture to be safe and submit for feedback. Unlike the final robot submission (the Veribot), the Betabot does not yet need to be verified, but should provide best-thought-out conjecture in order to give students a head-start on the Veribot.
This teaches students by experience that most CPS designs are more challenging than it appears at first glance.
Students learn to appreciate the value of formal verification by seeing first hand the quality difference between their Betabots and their ultimate verified Veribots.
Feedback on the Betabots also prevents students from wasting effort on models that have fundamental flaws.

The course culminates in a self-defined final course project, with which the students compete in the CPS V\&V Grand Prix course competition, presenting to a panel of about 12 experts in CPS who give them feedback from an industry perspective from organizations such as Siemens, Google, Bosch, Aptiv, Galois, Argo AI, Near Earth Autonomy, MathWorks, Toyota, NASA, Intel, GM who also award prizes.
The mix of substantial conceptual challenges and self-defined course projects with written reports and slide presentations in the competition for the industry judges gives students a chance to shine and the instructor and judges a chance to get to know them fairly well.
Because of the time pressure and random effects of presentations at the competition, as well as mix of undergraduates, master's students, and PhD students competing, the final rank in the Grand Prix is no direct indicator of the overall skill of a student. In fact, the more detailed study of their course projects by my TAs and me frequently shines a complementary and more thorough light on their innovation and technical quality that they did not communicate perfectly to the judges.
But this is overall a much-appreciated factor of the LFCPS course that it provides so many different ways to shine and become a verification rock-star.

\paragraph{Evaluation.}
While the course evaluations are essentially perfect (with the exception of understandable concerns about the high workload), a much more useful indicator of success is the fact that so many students take this course, despite the fact that it is much easier to earn a good grade in other courses, and even though, when run with all its components, the course has a very intensive workload.
The author is since experimenting with ways of reducing the workload without over-proportionally reducing the learning outcomes.

Anecdotal evidence since the introduction of active learning quizzes into the LFCPS course led to students who are significantly advanced compared to their peers from prior years.
This experiment is imperfect, however, because active learning quizzes were introduced at the start of the covid-19 pandemic, where the course became fully remote, so that more than one aspect of the course was changed at the same time.
Statistical analysis also indicates that the active learning quiz scores are the strongest predictor for the overall course grade.

\section{Course: Programming Language Semantics}

The Programming Language Semantics course that the author taught at Carnegie Mellon University was a logical redesign benefitting significantly from previous course designs by Steve Brookes and John Reynolds \cite{Reynolds}.
The use of logic in studying programming language semantics is universal wisdom, most obviously in the case of axiomatic semantics for programming languages, which establishes logical axioms that characterize the truth of statements about programs.
Other parts of programming language semantics courses deal with several variations of denotational and operational semantics.
The relation between those different flavors of semantics is most exciting and valuable.
The equivalence of denotational and operational semantics, which helps combine the advantages of both, and the soundness and relative completeness relations of denotational or operational and axiomatic semantics, teach valuable lessons about how to justify that different perspectives on a programming language are in sync.
But different designs of programming language semantics courses differ in how pervasively logic is used.
Tobias Nipkow, for example, propagates the pervasive use of the theorem provers Isabelle/HOL to study programming language semantics \cite{DBLP:conf/vmcai/Nipkow12,DBLP:books/sp/NipkowK14}.
The author's Programming Language Semantics (PLS) course does not use formal theorem provers, which reduces the learning curve of proof tools, but still consistently benefits from the use of dynamic logic and uniform substitution \cite{DBLP:journals/jar/Platzer17}.

\paragraph{Partial Semantics.}
One obvious difference impacted by the use of logic is in the development of program states.
For PLS courses that are developed from the perspective of programming languages, it is natural to define states as partial functions $\omega:\mathcal{V}\dashrightarrow\mathcal{D}$ from variables in $\mathcal{V}$ to values in $\mathcal{D}$ and then define the meaning of expressions and programs and assertions about programs equally partially while distinguishing cases on whether or not a state is defined on all variables that a program reads or that a formula depends on.
For instance, the semantics $\omega\lenvelope x+1\renvelope$ of a program expression $x+1$ in the state $\omega$ is only defined if the state $\omega$ actually gives a value to the variable $x$ that the expression $x+1$ mentions.
Otherwise, $\omega\lenvelope x+1\renvelope$ remains undefined.
This detail is useful to teach about delicacy and precision, and prepares for other reasons of undefinedness in program execution such as null-pointer dereferences or out of bounds array accesses, but it does cause a fair amount of technical hurdles and complications at every step along the way if every connection of syntax and semantics needs to be guarded by an assumption that the states are defined everywhere where they need to be to make sense of the syntax.

\paragraph{Impartial Semantics.}
For PLS courses developed from the perspective of logic, it is more natural to define states as total functions $\omega:\mathcal{V}\to\mathcal{D}$ from variables to values, which makes it easier to evaluate terms and formulas and programs as a function of the values of the variables they read.
The semantics $\omega\lenvelope x+1\renvelope$ is always defined and equas $\omega(x)+1$, since the total function $\omega$ has some value $\omega(x)$ for every variable $x\in\mathcal{V}$.
The resulting substantial conceptual simplicity is a benefit even if states ``waste information'' in the sense that variables have values even if they are never needed.
With the establishment of an easily proved coincidence lemma, saying that the value of formulas and the effect of programs only depends on the values of their free variables $\freevars{\cdot}$, this difference between both approaches vanishes, but the conceptual simplicity and absence of technicalities of undefinedness remains a benefit of the approach coming from logic.

\begin{lemma}[Coincidence lemma \cite{DBLP:journals/jar/Platzer17}]
  If $\omega=\nu$ on $\freevars{\theta}$, then \(\omega\lenvelope\theta\renvelope=\nu\lenvelope\theta\renvelope\).
\end{lemma}
After establishing this coincidence lemma it becomes clear retroactively that only the values of some part of the state affect the meaning of terms $\theta$ (and similarly for other expressions).
But undefinedness or the need to check for the semantic compatibility of states and expressions was never a concern.

\paragraph{Syntactic Transformation via Special-purpose Semantics.}
Even more pronounced is the simplicity afforded by logic for program transformations and program contexts.
A natural operation on programs is to replace one part with something else.
Obviously, some program transformations change the program behavior while others do not.
Proving when program transformations do not affect the program behavior is surprisingly difficult without the use of logic even in simple cases.
To illustrate, a transformation turning a part $e$ of a program into $k$ conventionally needs the definition of a program context as a program $\alpha(\_)$ with a hole $\_$ that is once filled with the syntactic part $e$ to form $\alpha(e)$ and that is once filled with the syntactic part $k$ to form $\alpha(k)$ giving the following rewrite:
\[
\alpha(e) \leadsto \alpha(k)
\]
Besides defining the syntax of programs, the syntax of programs with holes, the syntactic operation of the program transformation, a correctness argument also needs to define not just the semantics $\lenvelope\alpha\renvelope_{\text{prg}}$ of programs $\alpha$ but also the semantics $\lenvelope\alpha(\_)\renvelope_{\text{cxt}}$ of programs $\alpha(\_)$ with holes, and the semantics $\lenvelope e\renvelope_{\text{part}}$ of the affected parts $e$.
And one then has to prove that the semantics of the program obtained after the syntactic transformation of plugging in $e$ into the program $\alpha(\_)$ with hole $\_$ to obtain the program $\alpha(e)$ is equivalent to the semantics $\lenvelope\alpha(\_)\renvelope_{\text{cxt}}$ of the program with hole $\alpha(\_)$ applied to the semantics $\lenvelope e\renvelope_{\text{part}}$ of the part $e$:
\begin{equation}
\lenvelope\alpha(e)\renvelope_{\text{prg}} = \lenvelope\alpha(\_)\renvelope_{\text{cxt}} \big(\lenvelope e\renvelope_{\text{part}}\big)
\label{eq:compatible-semantics-of-plugged-holes}
\end{equation}

Of course, inductively proving this particular compatibility result \rref{eq:compatible-semantics-of-plugged-holes} teaches one to scrutinize and relate several different but closely related versions of semantics definitions with inductions on the syntax and corresponding decompositions on the semantics.
But the proof of \rref{eq:compatible-semantics-of-plugged-holes} and similar results is fairly technical and all the required definitions repetitive even if the semantic domain shifts a little from program behavior to functions from hole-filling behavior to program behavior.
Likewise one will have to prove that the semantics of a program context $\alpha$ that does not, in fact, even have a hole is actually equivalent whether taken as a program or as context semantics:
\[
\lenvelope\alpha\renvelope_{\text{prg}} = \lenvelope\alpha\renvelope_{\text{cxt}}(S) 
\quad\text{for any possible semantics $S$ that a part $e$ may have}
\]

\paragraph{Syntactic Transformation via Logic.}
Taking logic seriously leads to another angle on PLS courses, where the use of dynamic logic and uniform substitution significantly simplify technical challenges.
Uniform substitution was originally defined for first-order logic by Church \cite[\S35,40]{Church_1956} for substituting function symbols with terms and substituting predicate symbols with formulas, uniformly everywhere, while respecting that free variables cannot be bound during the substitution \cite{DBLP:journals/jar/Platzer17}.
Corresponding generalizations lift uniform substitution to dynamic logic with programs, where, in addition, program symbols can be substituted with programs \cite{DBLP:journals/jar/Platzer17}.
The uniform substitution proof rule \irref{US} states that if a formula $\varphi$ has a proof then all its uniform substitution instances $\applyusubst{\sigma}{\varphi}$ have a proof.

\begin{theorem}[Soundness of uniform substitution \cite{DBLP:journals/jar/Platzer17}] \label{thm:usubst-sound}
  The proof rule \irref{US} is sound (where \irref{US} is only applicable if the substitution result $\applyusubst{\sigma}{\varphi}$ is defined).
\[
\cinferenceRule[US|US]{uniform substitution}
{\linferenceRule[formula]
  {\varphi}
  {\applyusubst{\sigma}{\varphi}}
}{}
\]
\end{theorem}
Proving soundness of the uniform substitution proof rule \irref{US} takes some effort as well but only needs to be established once and for all as the \emph{only form of syntactic transformation}.
A similar proof principle uses the same uniform substitution $\sigma$ simultaneously on all premises and the conclusion of a (locally sound) inference.

\begin{theorem}[Soundness of uniform substitution of rules \cite{DBLP:journals/jar/Platzer17}] \label{thm:usubst-rule}
  All uniform substitution instances (whose substitutions introduce no free variables) of locally sound inferences are locally sound:
  \[
\linfer
{\phi_1 \quad \dots \quad \phi_n}
{\psi}
~~\text{locally sound}\qquad\text{implies}\qquad
\linfer%
{\applyusubst{\sigma}{\phi_1} \quad \dots \quad \applyusubst{\sigma}{\phi_n}}
{\applyusubst{\sigma}{\psi}}
~~\text{locally sound}
\irlabel{USR|USR}
  \]
\end{theorem}
Correctness of the syntactic transformation of term $e$ for an equal term $k$ is then a uniform substitution instance of the following obvious axiomatic proof rule:
\[
      \cinferenceRule[CQ|CQ]{congequal congruence of equations on formulas (convert term congruence to formula congruence: term congruence on formulas)}
      {\linferenceRule[formula]
        {f = g}
        {p(f) \lbisubjunct p(g)}
      }{}%
\]
The congruence rule \irref{CQ} states that if (nullary function symbol) $f$ is equal to $g$ then for any (unary) predicate $p$ the formula $p(f)$ is equivalent to $p(g)$, which is evidently correct by the principle of substitution of equals for equals.
Making a concrete inference with rule \irref{CQ} simply amounts to using the uniform substitution principle from \rref{thm:usubst-rule} to substitute the concrete term $e$ for the function symbol $f$ and the concrete term $k$ for the function symbol $g$ and a concrete property of a concrete program for the predicate symbol $p$ in which, by way of the principle that replacements of predicate symbols will still have their arguments in the same places, implicitly defines a context without the need for separate definitions, semantics and constructions.
Thanks to uniform substitution, which modularizes all the challenges of how syntactic transformations preserve the semantics, independently of the particular use case, congruence reasoning and contextual replacements reduce to the self-evident rule \irref{CQ}.

The same principle of substituting equals for equals immediately transfers to the situation when replacing formulas for equivalents or when replacing subprograms without the need to define a new semantics and new syntactic replacement principles and their semantic compatibility from scratch.
For example, congruence on terms reduces to uniform substitution uses via \rref{thm:usubst-rule} of the evident axiomatic proof rule \irref{CT}:
\[
      \dinferenceRule[CT|CT]{congterm congruence on terms}
      {\linferenceRule[formula]
        {f = g}
        {c(f) = c(g)}
      }{}%
\]
No new syntactic category of terms with holes or new semantics or new replacement mechanisms or new compatibility proofs are needed.
Everything simply follows from uniform substitution.

\paragraph{Logical Compiler Optimizations.}
Likewise, program transformations such as common subexpression elimination are but uniform substitution instances of the backwards direction of the assignment axiom:
\[
\cinferenceRule[assignb|$\dibox{:=}$]{assignment / substitution axiom}
{\linferenceRule[equiv]
  {p(\genDJ{x})}
  {\axkey{\dbox{\pupdate{\umod{x}{\genDJ{x}}}}{p(x)}}}
}{}
\]
The idea is merely that a common subexpression $\genDJ{x}$ identified in a formula $p(\genDJ{x})$ (or a program contained therein for the purpose of showing equivalence of the transformation) that may occur multiple times in different places is pulled out and stored into a common variable $x$ that is assigned to after computing the value of the term $\genDJ{x}$ only once.

\begin{example}[Common subexpression elimination via logic] \label{ex:CSE}
The common subexpression $a^2+b$ can be pulled out of the following program
\[
\dbox{\pwhile{y^2<a^2+b}{\lpbrace z:=z+y^2*(a^2+b);y:=y+2*3\rpbrace}}{P}
\]
to obtain the, by \irref{assignb} equivalent
\[
\dbox{x:=a^2+b;
\pwhile{y^2<x}{\lpbrace z:=z+y^2*x;y:=y+2*3\rpbrace}}{P}
\]
But common subexpression elimination cannot pull out any of the $y^2$ occurrences, because $y$ is changing afterwards in the loop body which will be before the (textually earlier) next use of $y$ in the next round of the loop.
This difference of applicability of common subexpression elimination, which is crucial for correctness, is easily spotted by uniform substitution via \rref{thm:usubst-sound} applied to axiom \irref{assignb}.
Of course, another intuitive giveaway is that, if all parts of a loop condition were common subexpressions, then the loop condition never changes its truth value, so the loop either never runs or runs forever.
\end{example}

Ordinarily, special-case analyses would have to be designed and their correctness proven separately for syntactic transformations such as common subexpression elimination in programming language compilation.
The consequent use of logic reduces this to mere uniform substitution.

\begin{example}[Copy propagation via logic] \label{ex:copyprop}
Copy propagation is another compiler optimization that, of course, needs soundness-critical applicability checks.
Propagating the value of a variable to later occurrences of that variable is again just uniform substitution via \rref{thm:usubst-sound} for the \irref{assignb} axiom.
This, for example, rephrases the last program of \rref{ex:CSE} equivalently to
\[
\dbox{x:=a^2+b;
\pwhile{y^2<x}{\lpbrace z:=z+y^2*(a^2+b);y:=y+2*3\rpbrace}}{P}
\]
or to
\[
\dbox{x:=a^2+b;
\pwhile{y^2<a^2+b}{\lpbrace z:=z+y^2*(a^2+b);y:=y+2*3\rpbrace}}{P}
\]
Again, uniform substitution directly tells apart this correct use of copy propagation from an incorrect attempt to propagate the value $z+y^2*x$ for any occurrence of $z$ (such as after the program or in itself), because that would already have a different value due to the loop.
\end{example}

\begin{example}[Constant folding via logic] \label{ex:constfold}
The constant folding transformation is merely a use of uniform substitution for the congruence rule \irref{CQ}, e.g., to reduce that the multiplication $2*3$ is $6$ in a program, because $2*3=6$ is true.
This reasoning rephrases the last program of \rref{ex:CSE} equivalently to
\[
\dbox{x:=a^2+b;
\pwhile{y^2<x}{\lpbrace z:=z+y^2*x;y:=y+6\rpbrace}}{P}
\]
If, for some reason, one would like to commute $a^2+b$ to $b+a^2$  (maybe in order to enable more common subexpression eliminations in other parts of the program), then another use of uniform substitution for congruence rule \irref{CQ} uses the equation \(a^2+b=b+a^2\) to transform the above program equivalently:
\[
\dbox{x:=b+a^2;
\pwhile{y^2<x}{\lpbrace z:=z+y^2*x;y:=y+6\rpbrace}}{P}
\]
\end{example}

With slight generalizations of uniform substitution to program relations such as refinement \(\asprg\refines\bsprg\) and equivalence \(\asprg=\bsprg\) \cite{DBLP:conf/cade/PrebetP24}, one can state a congruence rule for making use of program equivalences in any context $\contextapp{C}{\text{\textvisiblespace}}$
\[
\cinferenceRule[CP|CP]{congequiv congruence of equivalences on programs}
{\linferenceRule[formula]
  {\ausprg = \busprg}
  {\contextapp{C}{\ausprg} \lbisubjunct \contextapp{C}{\busprg}}
}{}%
\]

\begin{example}[Loop unwinding via logic]
Unwinding one round of a loop to run before the loop is easily done via uniform substitution via \rref{thm:usubst-rule} on congruence rule \irref{CP}.
Based on the equivalence
\[
\pwhile{\ivr}{\asprg} = \pifs{\ivr}{\lpbrace\asprg;\pwhile{\ivr}{\asprg}\rpbrace}
\]
the last program of \rref{ex:constfold} is transformed via \irref{CP} to its equivalent:
\begin{align*}
[x:=b+a^2;
\pifs{y^2<x}{\lpbrace z:=z+y^2*x;y:=y+6;\\
\pwhile{y^2<x}{\lpbrace z:=z+y^2*x;y:=y+6\rpbrace}\rpbrace}
]{P}
\end{align*}
If interesting optimizations now were to become possible in the pulled out first iteration, subsequent combinations of logical transformations can continue.
Here, this might apply, e.g., if the initial value of $y$ is known to satisfy $y^2<b+a^2$ such that the if branching is unnecessary.
\end{example}

\section{Conclusion and Outlook}

Overall, symbolic logic plays and should play a significant role in scientific education.
Logic both leads to significant simplifications of otherwise challenging concepts and logic leads to identifying the inherent core essence of ideas otherwise lost among lots of peripheral aspects.
As the experience relayed in this paper demonstrates, there is a wide variety of courses that benefit from the inclusion of ideas from logic ranging from introductory courses over systems courses to logic and programming language courses.
Logic helps on a scale ranging from tool-free informal logic reasoning supported by practice with mere dynamic checking of program-expressed contracts all the way to full feature theorem provers.
Active learning quizzes with theorem provers fully blackboxed in its autograder have an outside impact on the understanding of students that seems well worth the nonnegligible time investment.
The development of pedagogically well-paced and insightful questions aligned with the learning goals of the course is, however, a massive undertaking that can only be amortized by reusing the quiz questions and autograding infrastructure over the years.

\renewcommand{\doi}[1]{doi: \href{https://doi.org/#1}{\nolinkurl{#1}}}
\bibliographystyle{splncs04}
\bibliography{platzer,bibliography}
\end{document}